\documentclass[aps,prl,reprint,groupedaddress]{revtex4-2}
\usepackage{graphicx,float}
\begin{document}


\title{Liapunov exponent distributions and maps for multiple parameter logistic equation. Application to DNA and RNA sequences}


\author{Miguel Martín-Landrove}
\email[]{landrove.martin@ucv.ve}
\thanks{}
\affiliation{Centre for Medical Visualization, National Institute for Bioengineering, Universidad Central de Venezuela, Caracas, Venezuela}
\affiliation{Centre for Molecular and Medical Physics, Faculty of Science, Universidad Central de Venezuela, Caracas, Venezuela}
\affiliation{Centro de Diagnóstico Docente Las Mercedes, Caracas, Venezuela}

\author{B.P. Embaid}
\affiliation{Magnetism Laboratory, Centre for Condensed Matter Physics, Faculty of Science, Universidad Central de Venezuela, Caracas, Venezuela}

\date{\today}

\begin{abstract}
The multiple-parameter logistic equation has previously been utilized to determine the global stability of ternary codes, based on the arrangement of different symbols within the code. This approach has been extended to DNA and RNA sequences, proposing a specific application in the context of reading and translation processes involved in DNA replication and RNA-mediated protein codification. To address the complexity of mapping Liapunov's exponents in terms of four parameters—representing the different nucleotide bases—specialized mapping techniques have been developed. These include Liapunov exponent distributions for entire sequences, as well as binary maps that classify nucleotide bases based on their chemical type (purinic or pyrimidinic). Such methodologies provide a framework for examining the structural and functional properties of genetic material. The sequences analyzed encompass a wide range of DNA and RNA types, including those with and without introns, as well as codifying and non-codifying regions. This multifaceted approach offers valuable insights into the dynamic behavior and stability of nucleotide arrangements, contributing to a deeper understanding of the underlying processes that govern genetic replication and protein synthesis.
\end{abstract}


\maketitle

\section{Introduction}
The study of DNA sequences is fundamental to understanding the molecular basis of life, providing insights into genetic information, structural properties, and functional mechanisms. Over the years, various mathematical and computational techniques have been employed to analyze DNA sequences, uncovering patterns and correlations that contribute to biological function and evolution. Among these techniques, dynamical systems theory offers a powerful framework for exploring the complexity and predictability inherent in genetic data.

Liapunov exponents, originally developed in the context of dynamical systems, are measures of the rate at which nearby trajectories diverge or converge in a given system \citep{wolf1985}. These exponents have found applications in diverse scientific disciplines, from fluid dynamics to financial systems, as they provide a quantitative measure of chaos and predictability. In the context of DNA analysis, Liapunov exponents offer a novel approach to characterizing sequence complexity and identifying potential regulatory patterns.

Recent studies have highlighted the potential of Liapunov exponents to reveal structural and functional aspects of biological sequences. They have been used to detect conserved motifs, distinguish coding from non-coding regions, and identify repetitive elements \citep{nicolis1989}. By treating DNA sequences as symbolic representations of dynamical systems, this approach allows researchers to leverage nonlinear dynamics to uncover hidden regularities in genomic data.

In particular, Liapunov exponents have been utilized to characterize the inherent stochasticity and deterministic behavior within DNA sequences \citep{peng1992,voss1992}. These approaches have demonstrated the feasibility of linking sequence features with biological functionality and evolutionary significance, emphasizing the utility of tools from nonlinear dynamics in genomics.

In this paper, we explore the application of Liapunov exponents to characterize DNA sequences systematically. We aim to investigate how this methodology can provide new insights into genomic organization and variability. Our work builds on previous efforts to apply dynamical systems theory in computational biology \citep{wolf1985} and extends the use of Liapunov exponents as a versatile tool for DNA sequence analysis.

\section{Materials and Methods}\label{sec1}

As initially proposed by Rössler et al. \citep{rossler1989} and Markus \citep{markus1990}, the logistic equation serves as a fundamental model in the study of nonlinear dynamical systems. Its application has been extended beyond traditional domains to analyze the stability and behavior of symbolic sequences, including those found in genetic material. The logistic equation,

\begin{equation}\label{eq:eq1}
	x_{t+1} = r_t x_t (1-x_t)
\end{equation}
can be studied in the case where $r_t$ can take multiple values according to a sequence. This extension introduces variability into the logistic model, aligning it more closely with complex biological systems. Notably, this idea has been linked to the potential application of the logistic equation in interpreting protein digestion processes mediated by specific enzymes. 

The same concept can be extended to the replication and protein codification (or synthesis) processes in DNA and RNA, which are also enzyme-mediated and typically occur sequentially. In these cases, it becomes necessary to consider more than two possible values for $r_t$, corresponding to the different nucleotide bases associated to the genetic code. This generalization allows the model to capture the inherent complexity of genetic sequences and the sequential enzymatic processes that drive replication and protein synthesis. 

Some modifications to the general scheme proposed by Rössler et al. \citep{rossler1989} and Markus \citep{markus1990} are necessary to develop useful maps for calculating the Liapunov exponents, particularly in the context of the general multiple-parameter case. One approach we explored was to calculate the global Liapunov exponent distribution by varying the entire set of parameters. However, this method proved to be less effective for scenarios where the focus is on altering the value associated with a single symbol while keeping the other parameters constant. This limitation highlights the need for more targeted strategies to isolate and analyze the effects of individual parameters on sequence stability and dynamics.

A general formula for calculating the Liapunov exponent, $\lambda$ for a sequence, often used in dynamical systems:

\begin{equation}
	\lambda = \lim_{n \rightarrow \infty} \frac{1}{n} \sum_{i=1}^n {\log_2 |f'(x_i)|}
\end{equation}
where $f'(x_i)$ is the derivative of the function $f(x)$ with respect to the state $x_i$ at the i-th iteration. For a logistic map or multi-parameter logistic equations related to DNA/RNA, this formula could adapt as follows,

\begin{equation}\label{eq:eq3}
	\lambda = \lim_{n \rightarrow \infty} \frac{1}{n} \sum_{i=1}^n {\log_2 |r_i (1-2x_i)|}
\end{equation}
where $r_i$ corresponds to the parameter value associated with a specific nucleotide (e.g., A, T, G, or C), and $x_i$ is the state at the step $i$ of the sequence under study.

\subsection{Two-parameters logistic equation}

In the context of the genetic code, equation (\ref{eq:eq3}) can be simplified by grouping the nucleotide bases into two categories. This classification can be performed in two distinct ways:
\begin{itemize}
\item Chemical Nature: The bases can be classified as either purines (adenine [A] and guanine [G]) or pyrimidines (cytosine [C] and thymine [T]), based on their molecular structures.\\

\item Bond Strength: Alternatively, the classification can be made based on the number of hydrogen bonds they form in the DNA double helix. Adenine (A) and thymine (T) pair through two hydrogen bonds, while guanine (G) and cytosine (C) form a stronger bond with three hydrogen bonds \citep{gould1994}. This classification is particularly relevant for exploring potential variations in DNA sequences caused by tautomerism—proton quantum tunneling between nucleotide base pairs \citep{lowdin1963,slocombe2021,slocombe2022}. Such phenomena could increase the likelihood of mutations, providing a potential mechanism for genetic variability.
\end{itemize}
    
\begin{figure}[h!]
\centering
\includegraphics[width=\columnwidth]{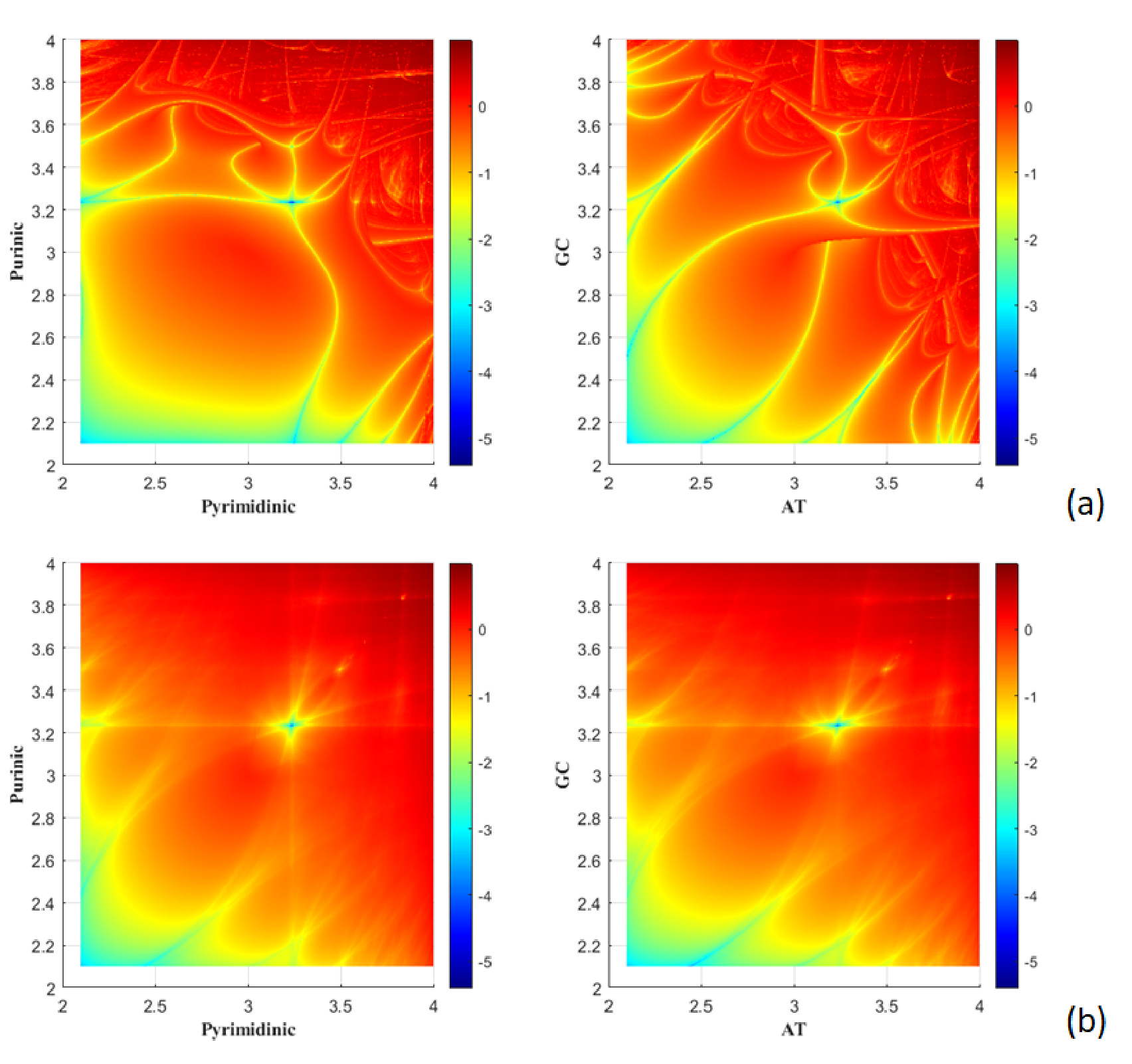}
\caption{Biparametric logistic series Liapunov exponent maps for different genes. On the left parameters are assigned according to chemical classification of nucleotide bases, i.e. pyrimidinic or purinic, and on the right, the parameters are assigned according to the strength of the nucleotide bases bond. (a) Human telomere sequence TTAGGG \citep{moyzis1991}, (b) Human c-jun proto oncogene (JUN).}
\label{fig:f2}
\end{figure}

An example of this approach is shown in Figure \ref{fig:f2}. The Lyapunov exponent maps reveal interesting patterns depending of the arrangement of the nucleotide sequences, in particular in Figure \ref{fig:f2}a, the DNA sequence is given by the repetition of the code TTAGGG while in Figure \ref{fig:f2}b, the DNA sequence corresponds to an almost random distribution of nucleotide bases.

\begin{figure}[h!]
\centering
\includegraphics[width=\columnwidth]{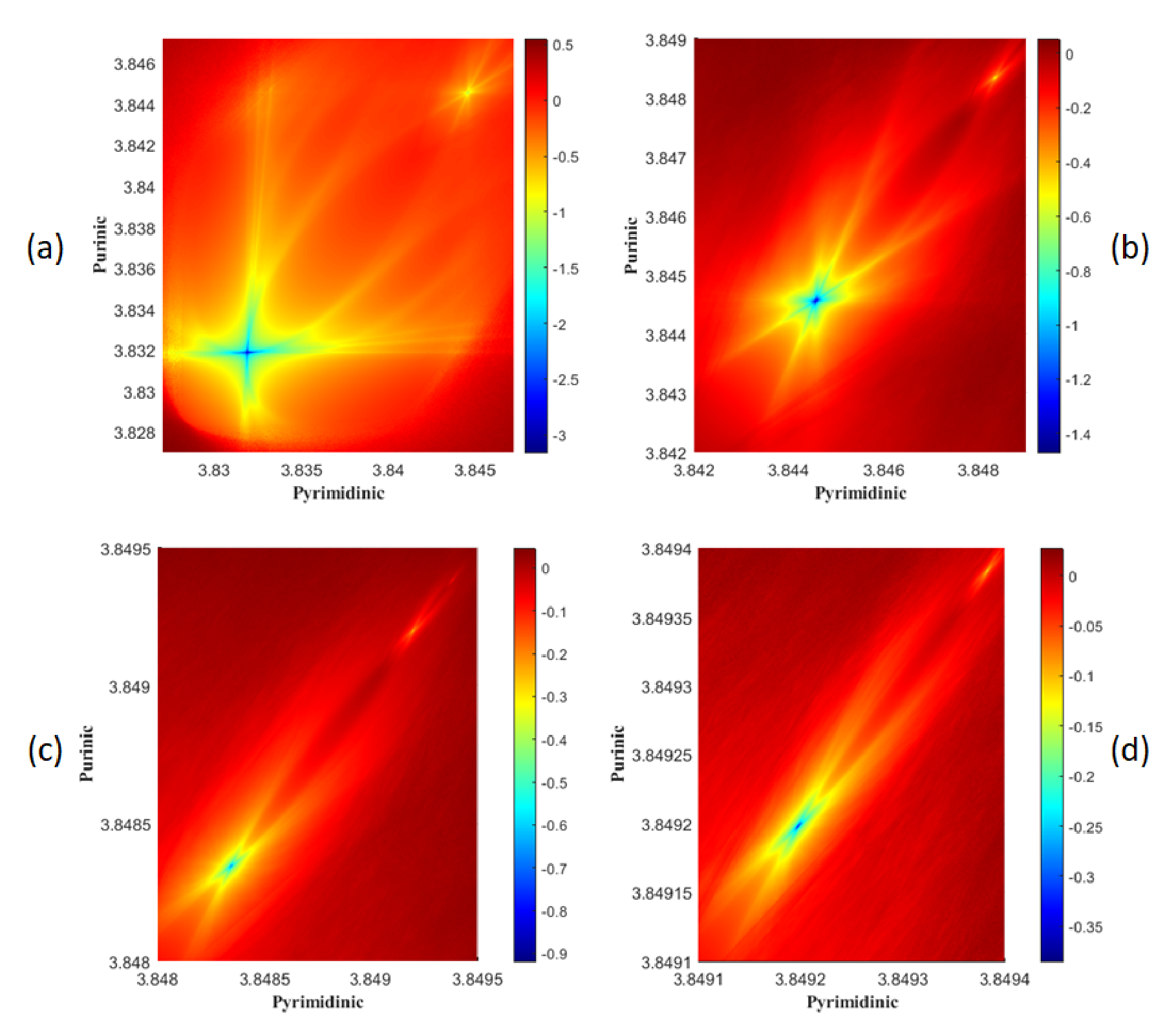}
\caption{Liapunov exponent maps for alpha-cardiac myosin heavy chain (MHC). The scale is reduced from (a) through (d).}
\label{fig:f3}
\end{figure}

Also, some patterns are reproduced at different scales as shown in Figure \ref{fig:f3}.

\subsection{Four-parameters logistic equation}

\begin{figure}[h!]
\centering
\includegraphics[width=\columnwidth]{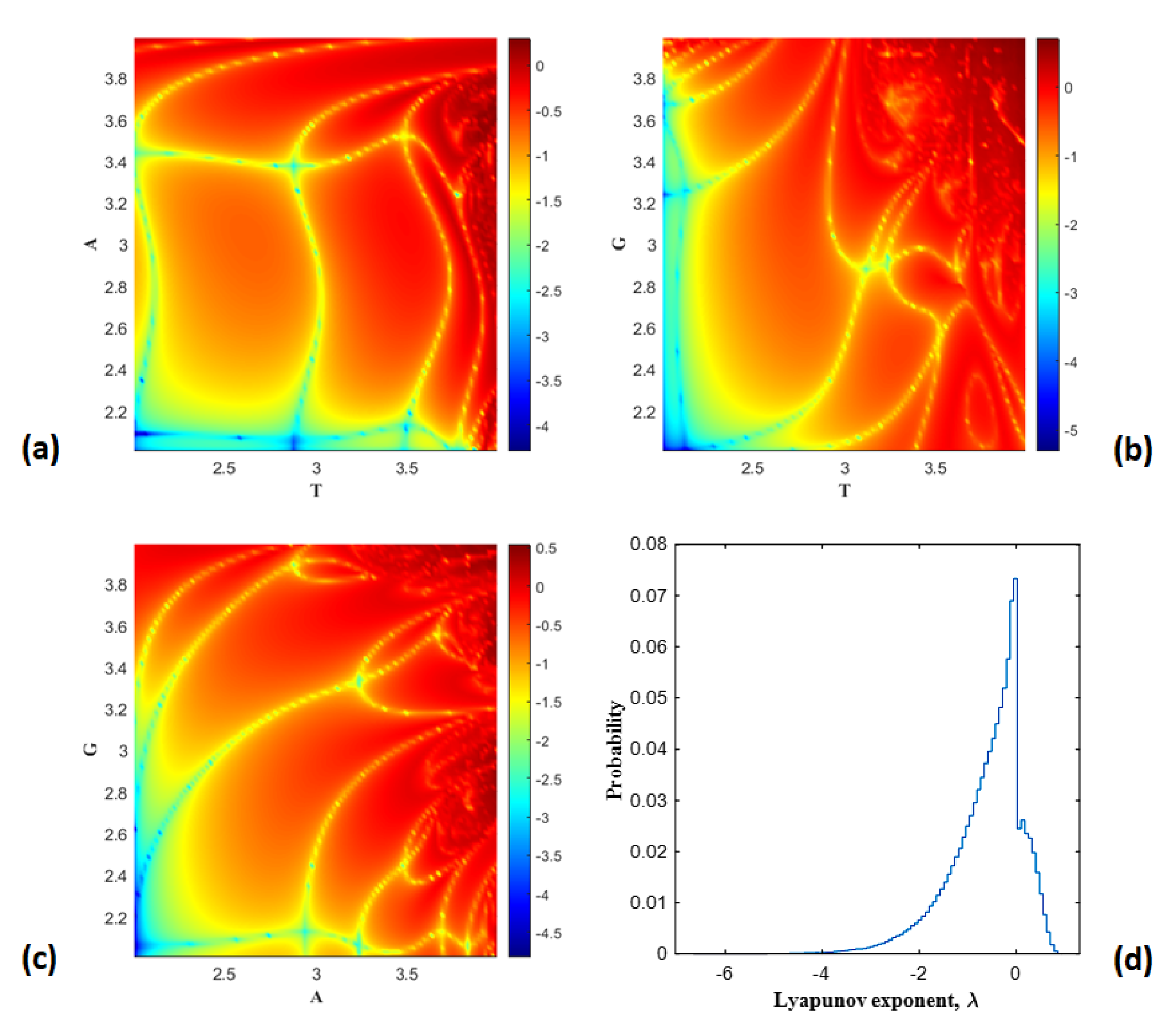}
\caption{Four-parameter logistic equation Lyapunov exponent maps for human telomere TTAGGG. For each map the parameter associated to certain bases was fixed to the value $2.4852$, (a) C-G, (b )A-C (c) T-C and (d) represents the Lyapunov exponent distribution}
\label{fig:f4}
\end{figure}

\section{Results and Discussion}

In Figure 1, a sequence of Liapunov exponent distributions is presented for the typical sequence found at the end of the human telomere, i.e., TTAGGGTTAGGG \citep{moyzis1991}, under different values of the fixed parameter. 

\begin{figure*}[ht]
\centering
\includegraphics[width=15cm]{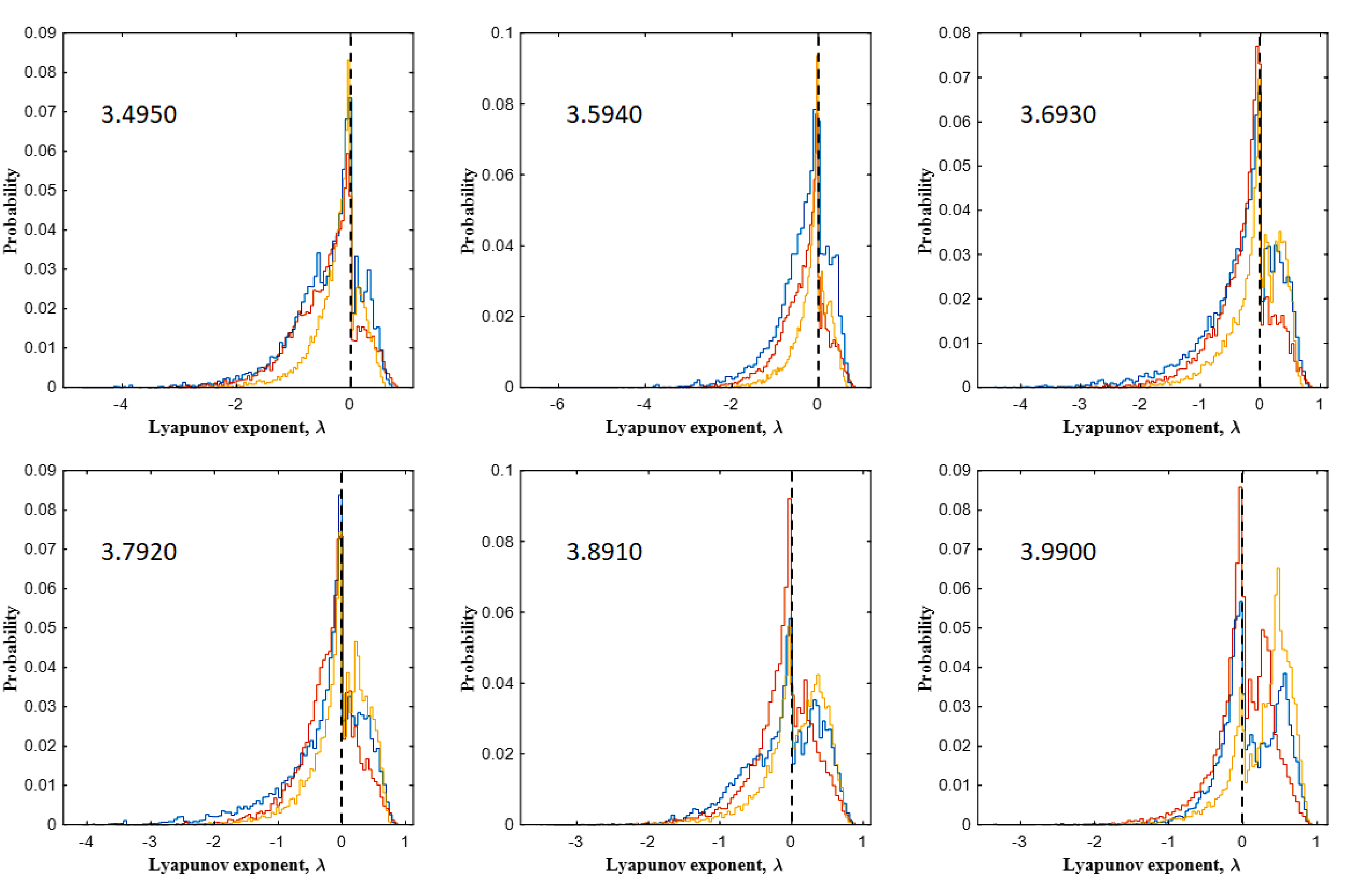}
\caption{Liapunov exponent distributions for human telomere TTAGGG. Different distributions correspond to keep one of the parameters fixed at the value shown in the figure, while varying the others. Colors indicate the fixed parameter, blue, T, red, A, and yellow, G.}
\label{fig:f1}
\end{figure*}

An alternative approach can be employed if it is possible to classify the symbols in the code based on specific properties, thereby reducing the dimensionality of the map. In the case of DNA and RNA, such classification can be performed based on the nucleotide base type, specifically as purinic (adenine and guanine) or pyrimidinic (cytosine and thymine/uracil) \citep{peng1992}. Using this approach, we obtained the results presented in Figure 2, which depict the Liapunov exponent distributions for the first 300 base pairs across five different DNA sequences.

In Figure 3, the results of the stable region (ranging between 3.82 and 3.86 in the parameters) for the first 300 base pairs of the human alpha-cardiac myosin heavy chain (MHC) gene are presented. Additionally, the global Liapunov exponent distribution for the complete gene, comprising 2,366 base pairs, is also shown. These results illustrate the stability characteristics within this parameter range and provide a comparative view of the local versus global dynamics of the gene.

Finally, in Figure 4, the Liapunov exponent distributions for various genes, each analyzed up to 300 base pairs, are presented.

\section{Conclusions}

The multiple-parameter logistic equation serves as a valuable tool for understanding highly complex codes, such as those found in DNA and RNA. Liapunov exponent distributions provide a means to compare the stability of different codes and assess the significance of specific symbols within them. Further research is required to interpret the biological implications of the proposed maps and distributions, enabling deeper insights into their relevance to genetic processes.

\section{Acknowledgments}
The authors thank the anonymous reviewers for their valuable suggestions.

\end{document}